\documentclass[aps,prd,secnumarabic,nobibnotes,twocolumn,superscriptaddress]{revtex4-1}
\usepackage{amsfonts}
\usepackage{mathrsfs}
\usepackage{amsmath}
\usepackage{color}
\usepackage{natbib}
\usepackage{graphicx}
\usepackage{bm}
\usepackage{amssymb}
\usepackage{xspace}
\usepackage{epstopdf}
\usepackage{dcolumn}
\usepackage{multirow}
\usepackage[colorlinks=true, letterpaper=true, pdfstartview=FitV, linkcolor=blue, citecolor=blue, urlcolor=blue]{hyperref}
\usepackage{wrapfig}

\makeatletter

\newcommand{\Rmnum}[1]{\expandafter\@slowromancap\romannumeral #1@}
\makeatother

\begin{document}
\title{Symmetry-enforced ideal lantern-like phonons in ternary nitride Li$_6$WN$_4$}

\author{Xiaotian Wang}\thanks{These authors have contributed equally to this work.}
\address{School of Physical Science and Technology, Southwest University, Chongqing 400715, China.}

\author{Feng Zhou}\thanks{These authors have contributed equally to this work.}
\affiliation{School of Physical Science and Technology, Southwest University, Chongqing 400715, China.}

\author{Tie Yang}
\address{School of Physical Science and Technology, Southwest University, Chongqing 400715, China.}

\author{Minquan Kuang}
\address{School of Physical Science and Technology, Southwest University, Chongqing 400715, China.}

\author{Zhi-Ming Yu}\email{zhiming$_$yu@bit.edu.cn}
\address{Key Lab of Advanced Optoelectronic Quantum Architecture and Measurement (MOE), Beijing Key Lab of Nanophotonics $\&$ Ultrafine Optoelectronic Systems, and School of Physics, Beijing Institute of Technology, Beijing 100081, China.}

\author{Gang Zhang}\email{zhangg@ihpc.a-star.edu.sg}
\address{Institute of High Performance Computing, Agency for Science, Technology and Research (A*STAR), 138632, Singapore.}

\begin{abstract}
Condensed matter systems contain both fermionic and bosonic quasiparticles. Owing to the constraint imposed by the Fermi level, ideal material candidate for the emergent particles with higher-dimensional degeneracy manifold (i.e., nodal lines and nodal surfaces) has not been found in electronic systems. This paper demonstrates that according to the first-principle calculations and symmetry analysis, realistic ternary nitride Li$_6$WN$_4$ features ideal (nearly flat) nodal-surface and nodal-line structures in its phonon spectra. These nodal degeneracies are shaped like lanterns, and their existence is guaranteed by nonsymmorphic symmetry. The corresponding topological phonon surface state covers exactly half the surface Brillouin zone (BZ) and can thereby be distinguished from those of conventional nodal-line and nodal-surface semimetals. The results of our study demonstrate the existence of ideal lantern-like phonons in realistic materials, which enriches the classification of topological quantum phases and provides a good basis for investigating the interaction between nodal-line and nodal-surface phonons in a single material.
\end{abstract}
\maketitle

Condensed matter systems~\cite{add1,add2} contain both fermionic and bosonic quasiparticles. The existence of the former in electronic structures~\cite{add3,add4,add5,add6,add7,add8,add8a} has been widely predicted and verified; by contrast, the existence of topological bosonic excitations~\cite{add9,add10,add11,add12,add13} has been rarely reported. According to the dimension of the degeneracy manifold, topological semimetals can be classified into nodal-point~\cite{add14,add15,add16,add17,add18,add19,add20}, nodal-line~\cite{add21,add22,add23,add24,add25}, and nodal-surface semimetals~\cite{add26,add27,add28}. Both nodal-line and nodal-surface semimetals exhibit many intriguing phenomena~\cite{add28,add30,add31,add32}. For example, it has been predicted that the nodal lines have drumhead-like surface states that cover a finite region in the surface Brillouin zone (BZ) or torus surface states spanning over the entire BZ. Although the nodal surface cannot have an intrinsic Chern number, it can have an induced one~\cite{add33}, which can be any integer value, depending on the chiral particles in the BZ. To determine the properties of nodal lines or nodal surfaces, the line or surface should preferably have a relatively flat energy dispersion. In addition, in electronic systems, the line or surface should be close to the Fermi level. Therefore, ideal nodal semimetals (in particular, ideal nodal-surface semimetals) are still missing in electronic systems.

Phonons, which are the basic emergent kind of bosons in crystalline lattices, can also display nontrivial degeneracy in their spectra. Because the Pauli exclusion principle does not apply to phonons, it provides a feasible basis for investigating bosonic excitations in a wide frequency range without the rigorous constraint of the fermion level. Researchers are currently searching for ideal topological phases in phonon systems~\cite{add9,add13,add34,add35,add36,add37,add38,add39,add40,add41,add42,add43,add44,add45}. In particular, the existence of Weyl, Dirac, triple point, nodal line phonons in the phonon spectra of several materials has been predicted, and the existence in some of them has been experimentally confirmed~\cite{add38,add45}. However, the existence of nodal-surface phonons has not been reported.

\emph{A natural question is whether ideal nodal-surface phonons can exist in realistic materials.} In this paper, we answer this question affirmatively. \emph{For the first time}, based on first-principle calculations and symmetry analysis, we predict the existence of nodal-surface phonons in ternary nitridotungstate Li$_6$WN$_4$, which has been experimentally synthesized~\cite{add46} via the solid state reaction of lithium subnitride (Li$_3$N) with W in a nitrogen atmosphere. The Li$_6$WN$_4$ material has two nodal surfaces in the $k_x$= $\pi$ and $k_y$= $\pi$ planes, respectively, the existence of which is guaranteed by nonsymmorphic screw-rotational symmetry and time-reversal symmetry. Remarkably, the two nodal surfaces are ideal in that (i) the two bands forming the nodal surfaces are well separated from other bands and in that (ii) the nodal surface is approximately flat in energy, with energy variations of less than 0.25 THz (approximately 1.00 meV). These characteristics have not been observed for other proposed nodal surface materials. Moreover, the two bands that form the nodal surfaces become degenerate in the Z-A path of the $k_z$= $\pi$ planes, thereby leading to two symmetry-enforced nodal lines. Eventually, the two nodal surfaces and two nodal lines in Li$_6$WN$_4$ result in exotic nodal-lantern phonon excitations, as illustrated in Fig. \ref{fig2}(c) and Fig. \ref{figs1}. Hence, this work presents novel topological phases of phonon systems,  predicts ideal material candidate, and paves the way for investigating nodal-lantern phonons in experiments.

The density-functional theory~\cite{add47} was used to calculate the ground state of this material, and the GGA-PBE formalism~\cite{add48} was used for the exchange-correlation functional. In addition, the projector augmented-wave method was used for the interactions between ions and valence electrons, and the energy cutoff was set to 600 eV. A $\Gamma$-centered $k$-mesh of 5$\times$5$\times$6 size was used to sample the BZ. Moreover, lattice dynamic calculations were performed to obtain the phonon dispersion of Li$_6$WN$_4$ at equilibrium lattice constants in the PHONOPY package~\cite{add49} with density-functional perturbation theory. The topological characteristics of the [001] phonon surface states were calculated by constructing a Wannier tight-binding Hamiltonian for phonons~\cite{add50}.
\begin{figure}
\includegraphics[width=8.8cm]{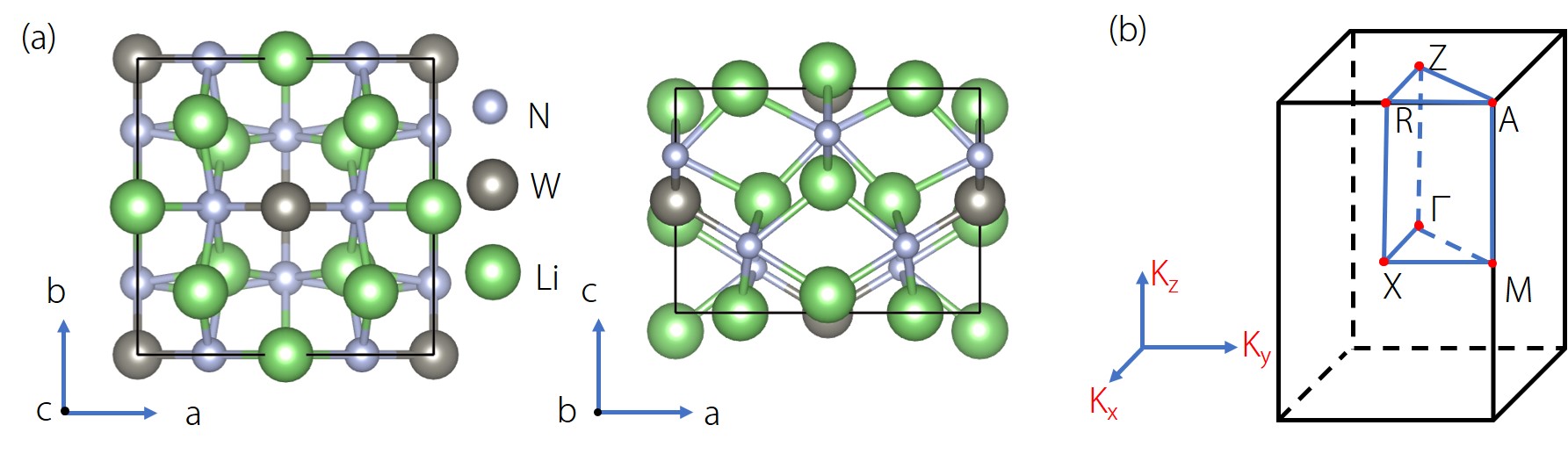}
\caption{Crystal structure of Li$_6$WN$_4$ material with P42/nmc type structure under different viewsides. (b) Bulk BZ and high-symmetry points.
\label{fig1}}
\end{figure}

As shown in Fig.~\ref{fig1}a, Li$_6$WN$_4$ is a tetragonal material with space group (SG) No. 137 (P42/nmc). The atomic positions are as follows: W is located at the 2a (0.25, 0.75, 0.25), N at the 8g (0.25, 0.995, 0.05), Li (1) at the 4d (0.25, 0.25, 0.338), and Li (2) at the 8f (0.537, 0.463, 0.25) Wyckoff positions. The crystal structure of Li$_6$WN$_4$ is completely relaxed according to the first-principle calculations; the computed lattice constants a = b = 6.71 {\r{A}} and c = 4.94 {\r{A}} agree well with the experimental results (a = b = 6.675 {\r{A}} and c = 4.928 {\r{A}}).

The symmetry operators of Li$_6$WN$_4$ are generated via fourfold screw rotation $S_{4z}$=$\{$${C_{4z}|00\frac{1}{2}}$$\}$, twofold screw rotation $S_{2x}$ =$\{$${C_{2x}|\frac{1}{2}\frac{1}{2}0}$$\}$, spatial inversion $\{$${I|\frac{1}{2}\frac{1}{2}0}$$\}$, and time-reversal symmetry ${\cal{T}}$.  Hence, Li$_6$WN$_4$ has three mirror symmetries: $M_x$, $M_y$, and $M_z$. Because we are interested in the phonon spectrum of Li$_6$WN$_4$, the spin-orbit coupling effect is neglected, indicating that the 2$\pi$ rotation equals 1 and ${\cal{T}}^2=1$.

In this work, the topological signature of the phonon dispersion of Li$_6$WN$_4$ was studied. The calculated phonon band along the high-symmetry paths (see Fig.~\ref{fig1}b) is shown in Fig.~\ref{fig2}a. The absence of imaginary frequency modes in the phonon dispersion indicates that Li$_6$WN$_4$ is dynamically stable. We focused on the two phonon bands appearing in the 22.5-24.5 THz range; they are well separated from other energy bands. Evidently, the two phonon bands become twofold degenerate along the X-M-A-R-X and Z-A paths. For clarity, we divided the doubly degenerate phonon bands in the 22.5-24.5 THz range into the two regions R2 and R3, as shown in Fig.~\ref{fig2}b. The degeneracies in the R2 and R3 regions are discussed separately.

First, the double degeneracies along the X-M-A-R-X path were studied (R2 in Fig.~\ref{fig2}b). All these high-symmetry paths lie in the BZ boundary, e.g., in the $k_x$= $\pi$ plane (see Fig.~\ref{fig2}c). In Fig.~\ref{fig3}a and b, we divided the A-X paths into 5 parts and selected some other symmetry points along the A-X path: b1, b2, b3, and b4. The phonon dispersion along the a1-b1-a1$^{'}$, a2-b2-a2$^{'}$, a3-b3-a3$^{'}$, and a4-b4-a4$^{'}$ paths are shown in Fig.~\ref{fig3}c and Fig. S2, respectively. Evidently, two non-degenerated phonon bands linearly cross at b1, b2, b3, and b4.

In fact, the two phonon bands become degenerate in the entire $k_x$= $\pi$ plane (see the schematic diagram in Fig.~\ref{fig3}a), which leads to the creation of a nodal surface. The nodal surface is guaranteed by nonsymmorphic $S_{2x}$ symmetry and ${\cal{T}}$ symmetry because any generic point in the $k_x$= $\pi$ plane is invariant under ${\cal{T}}S_{2x}$ symmetry and ${({\cal{T}}S_{2x})}^2=-1$ in the $k_x$=$\pi$ plane. Owing to the fourfold screw rotation $S_{4z}$, there must exist another nodal surface in the $k_y$=$\pi$ plane. It should be noted that while the z-direction represents a screw rotation axis, there is no nodal surface in the $k_z$=$\pi$ plane, which can be directly inferred from the phonon dispersion along the R-Z path. This is because $S_{4z}^2$ is equivalent to a symmorphic operator and ${({\cal{T}}S_{4z}^2)}^2=1$ in the $k_z$=$\pi$ plane. However, although the existence of two nodal surfaces in the $k_x$=$\pi$ and $k_y$=$\pi$ planes is guaranteed owing to the symmetry characteristics, the energy dispersion of the surface is not constrained by these symmetries.

\begin{figure}
\includegraphics[width=8.0cm]{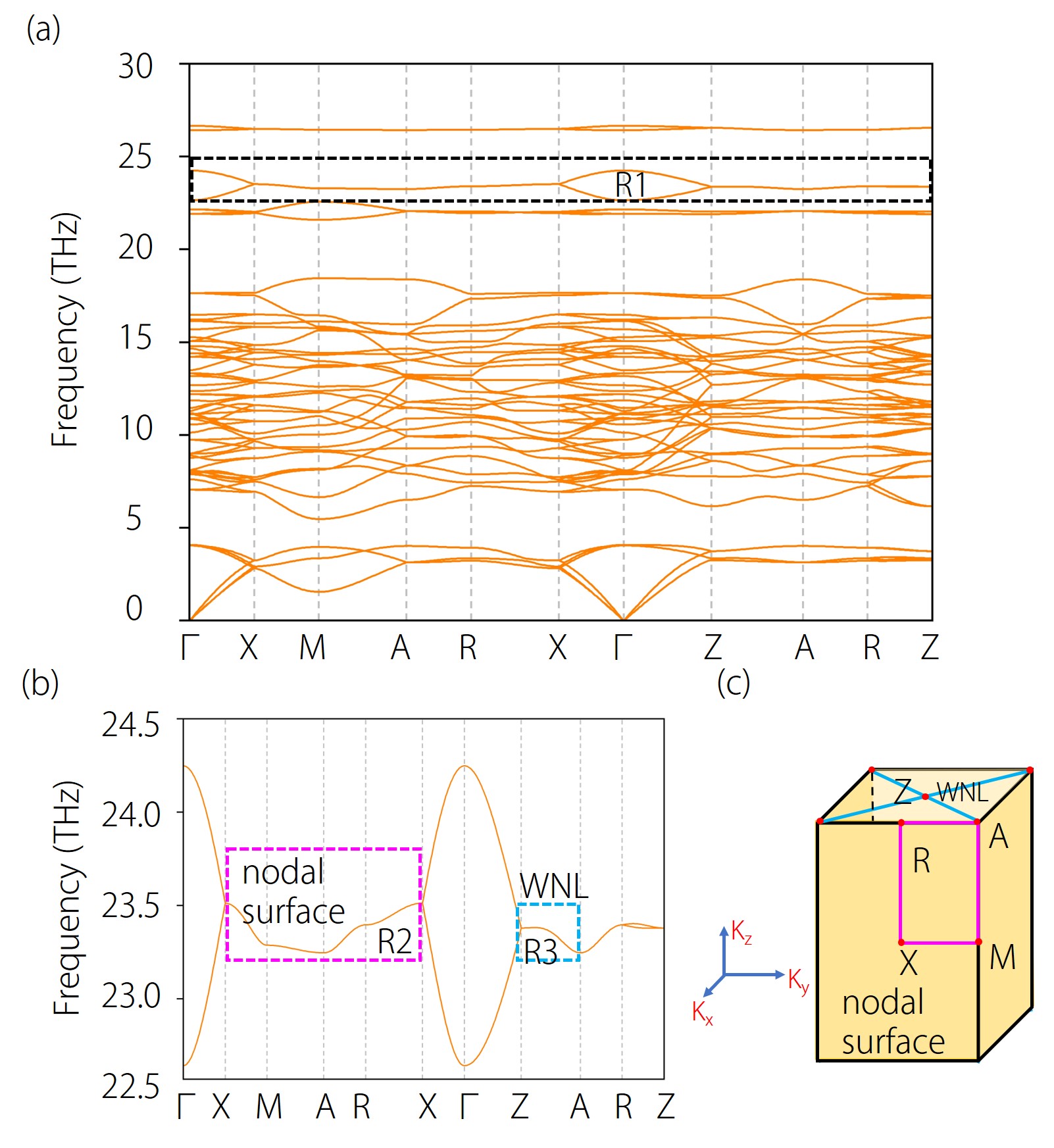}
\caption{ (a) Phonon dispersion of Li$_6$WN$_4$ along $\Gamma$-X-M-A-R-X-$\Gamma$-Z-A-R-Z paths; (b) enlarged frequencies of Li$_6$WN$_4$ with nodal surface phonons (see R2) and Weyl nodal line (WNL)-phonons (See R3);(c) Schematic diagram of nodal phonon states of R2 and R3 in 3D BZ.
\label{fig2}}
\end{figure}

To the best of our knowledge, the existence of an ideal topological nodal-surface semimetal with a flat nodal surface state has not been reported~\cite{add26}. Fortunately, the nodal-surface phonons in Li$_6$WN$_4$ are very flat in energy with energy variations of less than 0.25 THz (approximately 1.00 meV). More importantly, the phonon band structure is ``clean" in the sense that the bands forming the surface are well separated from the other bands. Another feature of the nodal-surface phonon is that it exhibits linear dispersion along the direction normal to the surface; in addition, the linear energy range is considerable large for most points on the surface, as indicated in Fig.~\ref{fig2}b. To present these characteristics more clearly, the calculated energy dispersion for points on the surface along the transverse direction is presented in Fig.~\ref{fig3}c and Fig. \ref{figs2}c, showing the linear energy range for the path away from the surface is considerable large.
\begin{figure}
\includegraphics[width=8.0cm]{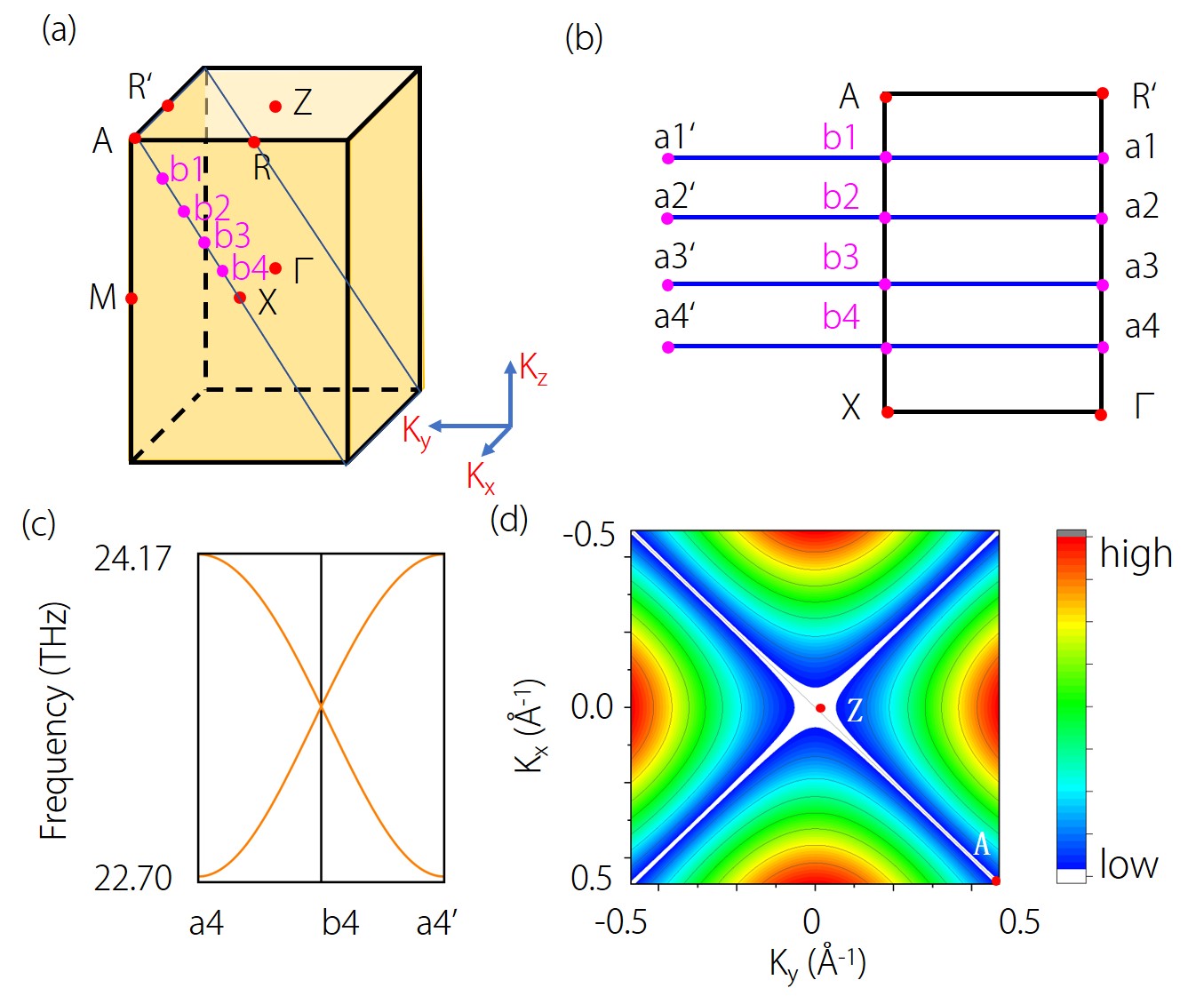}
\caption{ (a) Schematic diagram of nodal surface states in $k_x$=$\pi$ and $k_y$=$\pi$ planes; (b) some symmetry points of [011] plane; a1, a2, a3, and a4 and b1, b2, b3, and b4 are located at equal distances between R$^{'}$(A) and $\Gamma$(X); (c) phonon dispersion along a4-b4-a4$^{'}$ paths; (d) shape of one pair of Z-centered WNLs in $k_z$=$\pi$ plane.
\label{fig3}}
\end{figure}

In the next step, the degeneracies along the Z-A path are discussed (see R3 in Fig.~\ref{fig2}b). This is an essential nodal line protected by $M_z$ symmetry and twofold rotation along the (110) direction. The low-energy effective Hamiltonian expanding around a generic point on the Z-A path can be written as follows~\cite{add51}: \begin{equation}\label{FNRm}
\mathcal{H}=c_1+c_2 \sigma_3 (k_x-k_y)+c_3 \sigma_1 k_z,
\end{equation}
where $\sigma_i$ (i = 1, 2, 3) is the Pauli matrix, and the model parameter $c_i$ depends on the material and $k_{x(y)}$; the momentum is measured from the generic point on the Z-A path. To examine the nontrivial topological behavior of this nodal line, we calculate the Berry phase for a closed loop surrounding the line,  expressed as:
\begin{equation}\label{FNRm}
P_{B}=\oint_{c} \textbf{\emph{A}}(\textbf{\emph{k}}) \cdot d \textbf{\emph{k}}
\end{equation}
with $\textbf{\emph{A}}(\textbf{\emph{k}})=-i\left\langle\varphi(\textbf{\emph{k}})\left|\nabla_{\textbf{\emph{k}}}\right| \varphi(\textbf{\emph{k}})\right\rangle$  the Berry connection and $\varphi(\textbf{\emph{k}})$  the periodic part of the Bloch function. According to our results, we obtain $P_{B}=\pi$.  Then one knows that the nodal line along the Z-A path is topologically nontrivial, and  should result in interesting surface states. As previously mentioned, owing to the $S_{4z}$ symmetry, there are two nodal lines along the diagonal and clinodiagonal directions, respectively (see Fig.~\ref{fig2}c). The shape of the nodal lines in the $k_z$=$\pi$ plane is shown in Fig.~\ref{fig3}d, where one pair of nodal lines can be clearly observed (see the white lines). According to Fig.~\ref{fig2}b, the nodal-line phonon is relatively flat in energy, and its energy variation is less than 0.3 THz (approximately 1.24 meV).

Remarkably, the ideal nodal-surface phonon and ideal nodal-line phonon constitute a hitherto unobserved type of bosonic excitation, namely, a nodal-lantern phonon, as illustrated in Fig. \ref{fig2}(c) and Fig. \ref{figs1}. Furthermore, as the existence of the two nodal surfaces and two nodal lines is guaranteed by symmetry, any 3D material with SG 137 must have nodal-lantern phonons as long as two of its phonon bands are separated from the others.
\begin{figure}
\includegraphics[width=7.5cm]{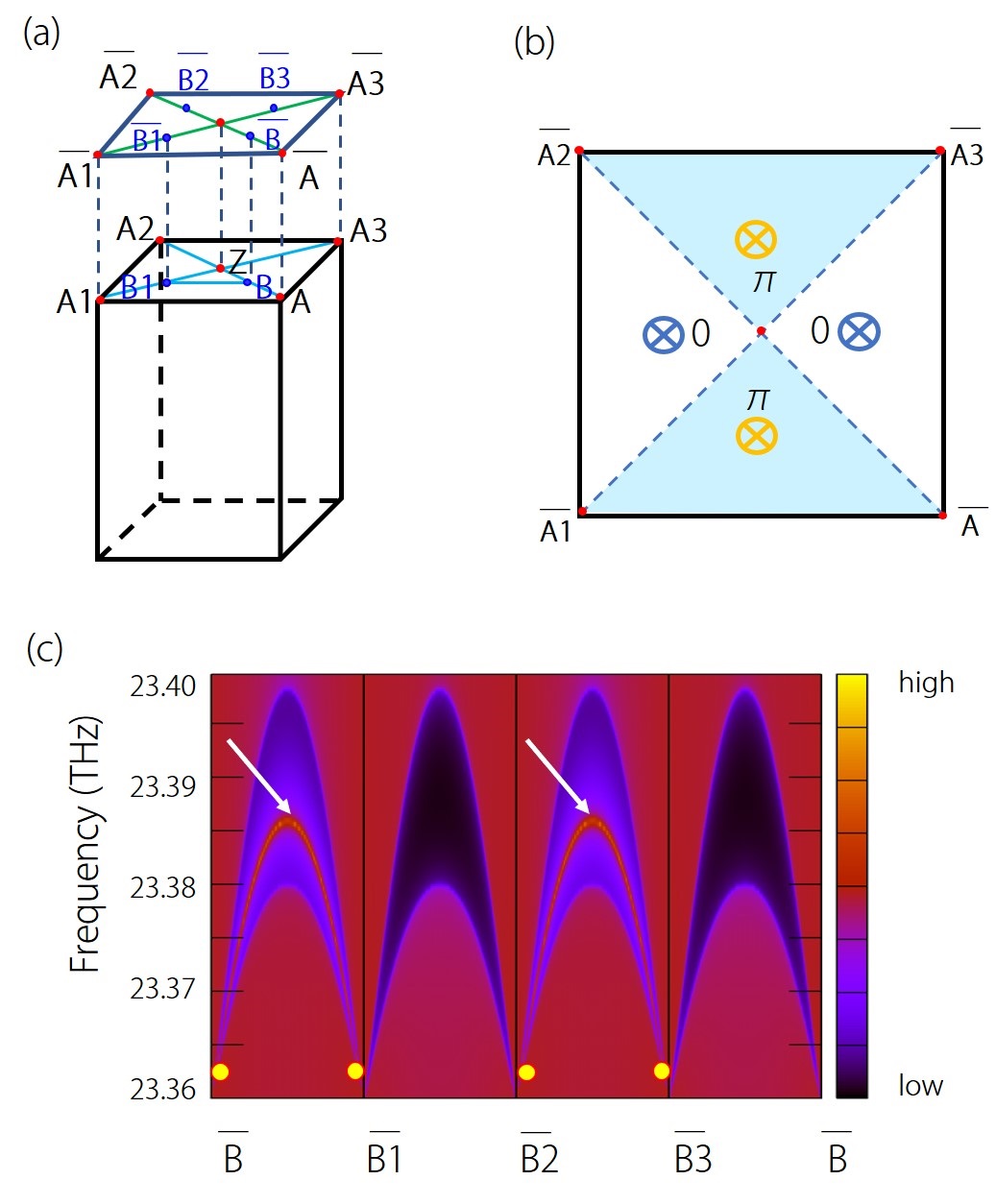}
\caption{(a) projection of WNL onto [001] surface and some surface symmetry points $\bar{A}, \bar{A}_{1}, \bar{A}_{2}, \bar{A}_{3}, \bar{B}, \bar{B}_{1}, \bar{B}_{2}, \bar{B}_{3}$ of [001] surface; (b) schematic diagram of surface state of [001] surface. Values 0 and $\pi$ are Zak phases for lines normal to [001] surface.(c) [001] phonon surface states of $\mathrm{Li}_{6} \mathrm{WN}_{4}$ along $\bar{B}-\bar{B}_{1} $ and $\bar{B}_{2}-\bar{B}_{3}$ surface paths;
\label{fig4}}
\end{figure}

In the next step, the phonon surface states corresponding to the nodal-lantern phonon are studied. Because the two boundaries of the BZ (i.e., the $k_{x}=\pi$ and $k_{y}=\pi$ planes) are covered by the nodal surface, the [001] surface is the only surface that can experience a clear surface state. The corresponding surface BZ and some surface symmetry points $\left(\bar{A}, \bar{A}_{1}, \bar{A}_{2}, \bar{A}_{3}, \bar{B}, \bar{B}_{1}, \bar{B}_{2}, \bar{B}_{3}\right)$ of the [001] surface are shown in Fig.~\ref{fig4}a. The calculated projected spectrum for the [001] surface along four surface paths $\bar{B}-\bar{B}_{1}$, $\bar{B}_{1}-\bar{B}_{2}$, $\bar{B}_{2}-\bar{B}_{3}$, and $\bar{B}_{3}-\bar{B}$ are shown in Fig.~\ref{fig4}c. Interestingly, evident [001] phonon surface states (highlighted by white arrows) can be found along the $\bar{B}-\bar{B}_{1}$ and $\bar{B}_{2}-\bar{B}_{3}$ surface paths. However, the phonon surface states have disappeared along the $\bar{B}_{1}-\bar{B}_{2}$ and $\bar{B}_{3}-\bar{B}$ paths. We use the Zak phase to explain this peculiar surface state. Here, the relevant Zak phase is the Berry phase along a straight line parallel to the $\Gamma$-Z direction and across the bulk BZ:
\begin{equation}\label{FNRm}
Z\left(k_{x}, k_{y}\right)=\sum_{n \in o c c} \oint_{c} A_{z}(\textbf{\emph{k}}) d k_{z},
\end{equation}
where $A_z$ is the z-component of the Berry connection. The summation is performed for all the occupied bands. Owing to $M_z$ symmetry, the Zak phase is quantized to 0 or $\pi$, which correspond to two topologically distinct phases. A $\pi$ Zak phase generally indicates the existence of nontrivial topological surface state~\cite{add52}. Because the two essential nodal lines along the Z-A path have nontrivial $\pi$ Berry phases, their projections onto the [001] surface divides the surface BZ into four regions. Two regions have $Z$=0, and two regions have $Z$=$\pi$, as illustrated in Fig.~\ref{fig4}b. Because each region occupies a quarter of the surface BZ, the topological surface state covers exactly half the surface BZ, which is consistent with the calculated results in Fig.~\ref{fig4}c.

Before closing the article, we would like to present some important remarks about the topological signatures of the Li$_6$WN$_4$ phonon dispersion: (i) Although the existence of nodal surface states in the fermionic electronic structures of realistic materials has been predicted~\cite{add26,add27,add28,add33}, the proposed nodal surface states of phonons have not been studied by other researchers. This paper presents the existence of ideal phononic nodal surface states in realistic materials \emph{for the first time}. (ii) The predicted nodal surface is ideal in that it is nearly flat in energy with energy variations of less than 0.25 THz; in addition, the two relevant phonon bands are the only ``clean" bands in the 22.5-24.5 THz range. (iii) The shape of the nodal structure in Li$_6$WN$_4$ (e.g., the nodal surface in the $k_x$=$\pi$ and $k_y$=$\pi$ planes and the nodal lines along the Z-A path) resembles that of a traditional lantern (see Fig. \ref{figs1}). We would like to point out that the nodal-lantern phonon is protected by symmetry and has never been reported before. These results can be used to investigate the entanglement between nodal-line and nodal-surface phonons.

In summary, based on first-principle calculations and symmetry analysis, a new topological phase was discovered: the nodal-lantern phonons in ternary nitride Li$_6$WN$_4$ are composed of two nodal-surface phonons in the $k_x$=$\pi$ and $k_y$=$\pi$ planes and one pair of nodal-line phonons in the $k_z$=$\pi$ plane. The existence of nodal-lantern phonons in Li$_6$WN$_4$ is protected by nonsymmorphic symmetries and time-reversal symmetry. Moreover, the phonon surface states in the [001] surface of this material were investigated; according to the results, the topological surface states of the nodal-lantern phonons cover exactly half the surface BZ. The presented results extend the concept of nodal surfaces to phonon systems, propose the existence of a undiscovered type of bosonic excitations, and predict the ideal material candidate.

\emph{\textcolor{blue}{Acknowledgments}} X.T. Wang is grateful for the support from the National Natural Science Foundation of China (No. 51801163) and the Natural Science Foundation of Chongqing (No. cstc2018jcyjA0765).

\end{document}